%% file: worstCaseDist.tex
\documentclass[times, 10pt,twocolumn]{article}
\usepackage{latex8}
\usepackage{times}

\usepackage{amsmath,epsfig}
\usepackage{amsfonts}
\usepackage{PageSetting}
\usepackage{wrapfig, epsfig, algorithm, algorithmic, graphicx}

\usepackage{times}

\def\done{\hspace*{\fill} $\framebox[1.5mm]{}$\smallskip}

\title{Anonymization with Worst-Case Distribution-Based Background Knowledge}

\author{%
{Raymond Chi-Wing Wong{\small $^1$}, Ada Wai-Chee Fu{\small $^2$}, Ke Wang{\small $^3$}, Yabo Xu{\small $^3$}, Jian Pei{\small $^3$}, Philip S. Yu{\small $^4$}}%
\vspace{1.6mm}\\
\fontsize{10}{10}\selectfont\itshape
$^1$The Hong Kong University of Science and Technology \hspace{2mm} \\
$^2$The Chinese University of Hong Kong \hspace{2mm}\\
$^3$Simon Fraser University \hspace{2mm} \\
$^4$University of Illinois at Chicago\\
{\fontsize{9}{9}\selectfont\ttfamily\upshape raywong@cse.ust.hk,
adafu@cse.cuhk.edu.hk},\\
{\fontsize{9}{9}\selectfont\ttfamily\upshape
\{wangk,yxu,jpei\}@cs.sfu.ca, psyu@cs.uic.edu}}

\begin{document}

\begin{sloppy}

\maketitle
\thispagestyle{empty}

\begin{abstract}
Background knowledge is an important factor in privacy preserving
data publishing.
Distribution-based background knowledge is one of the well studied
background knowledge. However, to the best of our knowledge, there
is no existing work considering the distribution-based background
knowledge in the worst case scenario, by which we mean that the
adversary has accurate knowledge about the distribution of sensitive
values according to some tuple attributes. Considering this worst
case scenario is essential because we cannot overlook any breaching
possibility. In this paper, we propose an algorithm to anonymize
dataset in order to protect individual privacy by considering this
background knowledge.
We prove that the anonymized datasets generated by our proposed
algorithm protects individual privacy. Our empirical studies show
that our method preserves high utility for the published data at the
same time.
\end{abstract}

\maketitle

\input{intro}

\input{probDef}


\input{probability}

\input{alg}
\input{exp}

\input{related}

\input{concl}

\bibliographystyle{latex8}
{
\small
\bibliography{worstCaseDist}
}

\input{proof}

\end{sloppy}

\end{document}

%% file: intro.tex

\section{Introduction}
\label{sec:intro}

Privacy preserving data publishing is an important topic in the
literature of privacy for very pragmatic reasons. As an example, AOL
did not take sufficient precaution and encountered some undesired
consequences. A dataset about search logs was published in 2006.
Later AOL realized that a \emph{single} 62 year old woman living in
Georgia can be re-identified from the search logs by some New York
Times reporters. The search logs were withdrawn and two employees
responsible for releasing the search logs were fired \cite{BJ06}.

\begin{table*}
\center  \scriptsize
\begin{tabular}{c c c}
\begin{minipage}[ht]{5.5cm}
 \begin{tabular}{ c | c | c | c |}\cline{2-4}
  Name & Nationality & Zipcode & Disease \\ \hline
  Alex & American & 55501 & Heart Disease \\ \cline{2-4}
  Bob & Japanese & 55502 & Flu\\ \cline{2-4}
   & Japanese & 55503 & Flu \\ \cline{2-4}
   & Japanese & 55504 & Stomach Virus\\ \cline{2-4}
   & French & 66601 & HIV\\ \cline{2-4}
   & Japanese & 66601 & Diabetes \\ \cline{2-4}
   & ... & ... & ... \\ \cline{2-4}
\end{tabular}
\caption{An example} \label{tab:rawData}
\end{minipage}
&
\begin{minipage}[ht]{4cm}
 \begin{tabular}{| c | c | c |}\hline
  Name & Nationality & Zipcode   \\ \hline
  Alex & American & 55501  \\ \hline
  Bob & Japanese & 55502 \\ \hline
  Chris & Japanese & 55503  \\ \hline
  David & Japanese &  55504 \\ \hline
  Emily & French & 66601 \\ \hline
  Fred & Japanese & 66601  \\ \hline
  ... & ... & ...  \\ \hline
\end{tabular}
\caption{Voter registration list} \label{tab:externalTable}
\end{minipage}
&
\begin{minipage}[ht]{7.2cm}
\begin{tabular}{c c}
\begin{minipage}[htbp]{3.8cm}
\center
\begin{tabular}{| c | c | c |}\hline
  Nationality &  Zipcode & GID\\ \hline
  American &  55501 & $L_1$ \\ \hline
  Japanese &  55502  & $L_1$ \\ \hline
  Japanese &  55503 & $L_2$ \\ \hline
  Japanese & 55504 & $L_2$ \\ \hline
  French &  66601  & $L_3$ \\ \hline
  Japanese & 66601 & $L_3$ \\ \hline
  ... & ... & ... \\ \hline
\end{tabular}
\end{minipage}
& \hspace*{0mm}
\begin{minipage}[htbp]{2.5cm}
\center
\begin{tabular}{| c | c |}\hline
  GID & Disease \\ \hline
  $L_1$ & Heart Disease \\ \hline
  $L_1$ & Flu\\ \hline
  $L_2$ & Flu \\ \hline
  $L_2$ & Stomach Virus\\ \hline
  $L_3$ & HIV\\ \hline
  $L_3$ & Diabetes\\ \hline
  ... & ...\\ \hline
\end{tabular}
\end{minipage}
\\
(a) QI Table &   (b) Sensitive table
\end{tabular}
\caption{A 2-diverse dataset anonymized from
Table~\ref{tab:rawData}} \label{tab:genTable} \label{tab4}
\end{minipage}
\end{tabular}
\end{table*}

\begin{example}[Data Publishing]\em
Suppose a table $T$ like Table~\ref{tab:rawData}
is to be anonymized for publication. Table
$T$ has two kinds of attributes, (1) the quasi-identifier (QI)
attributes and (2) the sensitive attribute. (1) The QI attributes
 can be used as an identifier
in the table.
In our example,  the QI attributes are Nationality and Zipcode.
Attribute Name is just for discussion and is not used for publication.
\cite{sweeney-kanonymity-model} points out that in a real dataset, about 87\% of
individuals can be uniquely identified by some QI attributes
with a publicly available external
table such as a voter registration list\footnote{There are many sources of such an external table.
Most municipalities sell population registers that include
the identifiers of individuals along with basic demographics;
examples include local census data, voter lists, city directories,
and information from motor vehicle agencies, tax assessors, and real
estate agencies \cite{samarati-protecting}. 
From
\cite{sweeney-kanonymity-model}, it is reported that a city's voter
list in two diskettes was purchased for twenty dollars, and was used
to re-identify medical records.}.
An example of a voter registration list is shown in
Table~\ref{tab:externalTable}.
(2) The sensitive
attribute contains some sensitive values.
In our example, the sensitive attribute is
``Disease" containing sensitive values such as Heart Disease and
HIV.
 Assume that each tuple in the table is owned by an
individual and each individual owns at most one tuple.

Our target is to anonymize $T$ and publish the anonymized dataset
$T^*$ like Table~\ref{tab4} to satisfy some privacy requirements.
A typical anonymization is described as follows. $T$ is horizontally
partitioned into multiple tuple groups. Let $P$ be a resulting
group. We give a unique ID called GID to $P$ and  all tuples in $P$
are said to have the same GID value. An anonymization defines a
function $\beta$ on each $P$ to form an \emph{anonymized group} (in
short, \emph{A-group}) such that the linkage between the QI
attributes and the sensitive attribute in the A-group is broken. One
way to break the linkage is
\emph{bucketization}, forming two tables, called the \emph{QI table}
(Table~\ref{tab:genTable}(a)) and the \emph{sensitive table}
(Table~\ref{tab:genTable}(b)): $P$ is projected on all QI attributes
and attribute GID to form the \emph{QI table}, and on the sensitive
attribute and attribute GID to form the \emph{sensitive table}.
 Therefore, a table $T$ is \emph{anonymized} to a dataset $T^*$ if
 $T^*$ is formed by first partitioning $T$ into a number
 of groups, then forming an A-group from
 each partition by $\beta$ and finally inserting each A-group into $T^*$.
%
For example, Table~\ref{tab:rawData} is anonymized to
Table~\ref{tab:genTable} by bucketization.
%
%
Such an anonymization is
commonly adopted in the literature of data publishing
\cite{XT06b,ZKS+07,MKM+07,WFW+07,LL08}.

There are many privacy models in the literature such as
$k$-anonymity \cite{sweeney-kanonymity-model}, $l$-diversity
\cite{l-diversity}, $t$-closeness \cite{LL07}, ($k, e$)-anonymity
\cite{ZKS+07}, Injector \cite{LL08} and $m$-confidentiality
\cite{WFW+07}. For illustration, let us consider a simplified
setting of the $l$-diversity model \cite{l-diversity} as a privacy
requirement for published data $T^*$. An A-group is said to be
\emph{$l$-diverse} or satisfy \textit{$l$-diversity} if in the
A-group the number of occurrences of any sensitive value is at most
$1/l$ of the group size. A table satisfies $l$-diversity (or it is
$l$-diverse) if all A-groups in it are $l$-diverse. 
Suppose that Table~\ref{tab:rawData} is anonymized to
Table~\ref{tab:genTable}.
Consider the A-group with GID equal to $L_1$ which corresponds
to the first two tuples in QI table (Table~\ref{tab:genTable}(a))
and the first two tuples in sensitive table (Table~\ref{tab:genTable}(b)).
In the following, we simply refer to the A-group with GID equal to $L_i$ by $L_i$.
Since $L_1$ contains two tuples, the group size of $L_1$ is equal to 2.
Since the number of occurrences of
any sensitive value (i.e., 1) is at most $1/2$ of the group
size, $L_1$ satisfies 2-diversity. Similarly, $L_2$ and $L_3$
satisfy 2-diversity.
Thus, Table~\ref{tab:genTable} satisfies 2-diversity.
The intention of 2-diversity is
that each individual cannot be linked to a disease with a
probability of more than 0.5 without any additional
background knowledge.

However, this table does not protect individual privacy sufficiently
if we consider background knowledge. \done
\end{example}

\begin{example}[Background Knowledge]\em
Consider $L_1$ in Table~\ref{tab:genTable}. In $L_1$, Heart Disease
and Flu are values of the sensitive attribute Disease. Since most
individuals can be re-identified by the QI attributes with a
publicly available external table such as voter registration list
\cite{sweeney-kanonymity-model}, if we are given the voter
registration list as shown in Table~\ref{tab:externalTable}, it is
easy to figure out that the two tuples in $L_1$ correspond to Alex
and Bob. From $L_1$, it \emph{seems} that each of the two
individuals, Alex and Bob, in this group has a 50\% chance of
linking to Heart Disease (Flu). The reason why the chance is
interpreted as 50\% is that the analysis is based on this group
without any additional information.

Suppose we are given a probability distribution as shown in
Table~\ref{tab:globalDistMotivatingExample}. 
The distribution of attribute set \{``Nationality"\} consists of the
probabilities that a Japanese, an American or a French is linked to
``Heart Disease" (and ``Not Heart Disease"). 
For example, the probability that
American is linked to Heart Disease is 0.1 and the probability that
Japanese is linked to Heart Disease is 0.003. With this
distribution, the adversary can say that Bob, being a Japanese, has
less chance of having Heart Disease. S/he can deduce that Alex,
being an American, has a higher chance of having Heart Disease. The
intended 50\% threshold is thus violated. \done
\end{example}

Hence background knowledge has important impact on privacy
preserving data publishing. Recent works
\cite{LL07,LLZ09,MKM+07,CLR07,WFW+07} start to focus on modeling
background knowledge.
\emph{Distribution-based background knowledge} is one type of the
well-known background knowledge which is used in the
state-of-the-art privacy model, $t$-closeness.
%
Distribution-based background knowledge \cite{LL07,LLZ09} is the
information related to the distribution about sensitive information
in data. There are at least two kinds of distribution-based
background knowledge, namely \emph{dataset based distribution} and
\emph{QI based distribution}. The dataset based distribution is
 the distribution of the values in the sensitive attribute according to the \emph{entire}
dataset \cite{LL07}. The QI based distribution is the distribution
of the values in the sensitive attribute restricted to individuals
with the same values on some QI attributes \cite{LLZ09}.

\begin{example}[Distribution-based background knowledge] \em
Suppose that there are 100,000 individuals in the dataset $T$ and
with 6,000 individuals linking to ``Heart Disease". The probability
that an individual $t$ in the
dataset is linked to ``Heart Disease" is 0.06. 
The dataset based distribution has been considered by \cite{LL07}.

In this paper we consider QI based distribution \cite{LLZ09}. Some
well-known examples of such knowledge are the facts that Japanese
seldom suffer from Heart Disease \cite{l-diversity} and male
individual cannot be linked to ovarian cancer \cite{LL08}. For
example, the distribution of the sensitive attribute according to
Japanese may be encoded as \{(Japanese:``Heart Disease", 0.003),
(Japanese:``Flu", 0.21), ...\} where (Japanese:$x$, $p$) denotes
that the probability that a Japanese is linked to a value $x$ is
$p$.
\done
\end{example}


%

\begin{table}[tb]
\center  \scriptsize
\begin{tabular}{c c c}
\begin{minipage}[ht]{8cm}
\center
\begin{tabular}{| c | c | c |} \hline
  $p()$ & Heart Disease & Not Heart Disease \\ \hline
  American & 0.1 & 0.9 \\
  Japanese & 0.003 & 0.997 \\
 French & 0.05 & 0.95 \\ \hline
\end{tabular}
\caption{A QI based distribution of attribute ``Nationality"
for our motivating example}
\label{tab:globalDistMotivatingExample}
\end{minipage}
\end{tabular}
\end{table}

If the QI based background knowledge is accurate, we say that we
have the worse case scenario. Considering the worst-case scenario is
essential in data publishing \cite{MKM+07,CLR07,WFW+07} because it
gives the maximal protection \cite{LL09}.
%
%
To the best of our knowledge, there is no existing
work considering the worst-case
QI based distribution.

There is only one work \cite{LLZ09} closely related to ours.
However, \cite{LLZ09} considers the QI based distribution background
knowledge with \emph{uncertainty}. Specifically, in \cite{LLZ09},
the uncertainty of the background knowledge is denoted by an input
parameter $B$.
\emph{Conceptually}, if $B$ is equal to 0, then the adversary has
the clearest understanding about background knowledge which
corresponds to the worst-case background knowledge. However, if $B$
is set to 0 in the model proposed by \cite{LLZ09}, then the
background knowledge is undefined. Also \cite{LLZ09} adopts a brute
force approach in the anonymization by checking the breaching
probability of anonymzied groups. There are two disadvantages on
this approach. The first problem is that the breaching probability
is hard to compute and therefore approximation is needed in their
method, which sacrifices the correctness. The second problem is that
the breaching probability is not monotone in that an A-group that
violates privacy may be split into two groups that preserve privacy.
Therefore, even though Mondrian \cite{multidimensional-Kanonymity} is
adopted as their algorithm, it does not guarantee an optimal
solution in spite of the effort in exhaustive search in each
iteration in the top-down processing. Our solution will overcome
 both of these problems.

Building on previous works, we propose a new method to handle the
worse case background knowledge. The essence of our method is the
following. We observe that privacy is breached whenever an
individual in an A-group has a much higher chance of linking to a
sensitive value compared with another individual in the A-group
according to the QI based distribution. Based on this observation,
we propose a solution which generates a dataset such that all
individuals in each A-group have ``similar" chances of linking to
any sensitive value in the group, according to the distribution.
For example if we form a group with an Americans and a Canadian,
linking to heart disease and flu, and suppose the probabilities of
Americans and Canadians being linked to heart disease and to flu are
similar. Since they have ``similar" chances, it is not possible for
the adversary to pinpoint any linkage of an individual to a
sensitive value with a higher chance. At the same time, our methods
can maintain high utility for the published table.

Our contributions can be summarized as follows. Firstly, to the best
of our knowledge, we are the first to handle the worst-case QI based
distribution. Secondly, we derive an interesting and useful
theoretical property and based on this property, we propose an
algorithm which generates a dataset protecting individual privacy in
the presence of the worst-case QI based distribution.
    Finally,
      we have conducted experiments
      which shows that our proposed algorithm is efficient and
      incurs low information loss.

%% file: probDef.tex
\section{Problem Definition}
\label{sec:probDef}

Let $T$ be a table. We assume that one of the attributes is a
sensitive attribute $X$ where some values of this attribute should
not be linkable to any individual. These values are called sensitive
values. The value of the sensitive attribute of a tuple $t$ is
denoted by $t.X$. A \textit{quasi-identifier} (QI) is a set of
attributes of $T$, $A_1, A_2, ..., A_q$, that may serve as
identifiers for some individuals. Each tuple in the table $T$ is
related to one individual and no two tuples are related to the same
individual. With publicly available voter registration lists (like
Table~\ref{tab:externalTable}), the QI values can often be used to
identify a unique individual
\cite{sweeney-kanonymity-model,WFW+07}.

There are two common approaches for anonymization, which generates
$T^*$ from $T$.
One is \emph{generalization} by generalizing all QI values in each
A-group to the same value. The other is \emph{bucketization}, which
we have illustrated in the previous section. For the ease of
illustration, we focus on bucketization. The discussion for
generalization is similar. 
%
With anonymization, there is a mapping which maps each tuple in $T$
to an A-group in $T^*$.
For example, the first tuple $t_1$ in Table~\ref{tab:rawData} is
mapped to A-group $L_1$.

The aim of privacy preserving data publishing is to deter any attack
from the adversary on linking an individual to a certain sensitive
value. Specifically, the data publisher would try to limit the
probability of
such a linkage that can be established. 

In the literature \cite{XT06b,WFW+07,LL08,LL07}, it is assumed that
the knowledge of an adversary includes (1) the published dataset
$T^*$,
(2) a publicly available external table $T^e$ such as a voter registration list that
maps QIs to individuals
\cite{sweeney-kanonymity-model,WFW+07} and (3) some background
knowledge. We also follow these assumptions in our analysis. We
focus on the QI based distribution as background knowledge.


The QI based distribution 
for the attribute set \{``Nationality"\} is described in
Table~\ref{tab:globalDistMotivatingExample}. 
%
Each probability in the table is called a \emph{global probability}.
The {\bf sample space} for each such discrete probability
distribution consists of the possible assignments of the sensitive
values such as $x$ to an individual with the particular nationality.
For nationality $s$, the sample space is denoted by $\Omega_s$.

Each possible value in attribute ``Nationality" in our example is
called a \emph{signature}. There are three possible signatures in
our example: ``Japanese", ``American" and ``French". In general,
there can be other attribute sets, such as \{``Nationality",
``Zipcode"\}, with their corresponding QI-based distributions. We
define the signature and the QI-based distribution for a particular
attribute set $\mathcal{A}$ as follows.

Given a QI attribute set
$\mathcal{A}$ with $q$ attributes $A_1, ..., A_q$. A
\emph{signature} $s$ of $\mathcal{A}$ is a set of attribute-value
pairs $(A_1, v_1),...,(A_q, v_q)$ which appear in the published
dataset $T^*$, where $A_i$ is a QI attribute and $v_i$ is a value. A
tuple $t$ in $T^*$ is said to \emph{match} $s$ if $t.A_i=v_i$ for
all $i = 1, 2, ..., q$.
%
For example, a signature $s$ can be \{(``Nationality", ``American"),
(``Zipcode", ``55501")\} if the attribute set $\mathcal{A}$
is \{``Nationality", ``Zipcode"\}. 
For convenience, we often drop the attribute names, and thus we have
\{``American", ``55501"\} for the above signature. The first tuple
in Table~\ref{tab:genTable}(a) matches \{``American"\} but the
second does not.

Given an attribute set $\mathcal{A}$, the \emph{QI-based
distribution} $G$ of $\mathcal{A}$ contains a set of entries $(s:x,
p)$ for each possible signature $s$ of $\mathcal{A}$, where $p$ is
equal to $p(s:x)$ which denotes the probability that a tuple
matching signature $s$ is linked to $x$.
%
For example, $G$ may contain (``Japanese":``Heart Disease", 0.003)
and (``American":``Heart Disease", 0.1). This involves two sample
spaces $\Omega_{Japanese}$ and $\Omega_{American}$.

%


\begin{definition}[$r$-robustness]
Given the QI-based distribution, a dataset $T^*$ is said to satisfy $r$-robustness
(or $T^*$ is $r$-robust) if, for any individual $t$ and any sensitive value $x$,
the probability that $t$ is linked to $x$, $p(t:x)$, does not exceed $1/r$.
\end{definition}

We will discuss about the sample space for $p(t:x)$ and derive a
formula for $p(t:x)$ in Section~\ref{sec:probabilityFormulation}. In
this paper, we are studying the following problem: given a dataset
$T$, generate an anonymized dataset $T^*$ from $T$ which satisfies
$r$-robustness and at the same time minimizing the information loss.
There have been different definitions for information loss in the
literature. In our experiments, we shall adopt the measurement of
accuracy in query results from $T^*$ versus that from $T$.


%% file: probability.tex
\begin{table*}
\center\small\scriptsize
\center 
 \scriptsize \small
\begin{tabular}{c c}
\begin{minipage}[ht]{2.5cm}
\begin{center}
\begin{tabular}{ | c | c |c |}\hline
  $p( )$ & $x$ &  $y$ \\ \hline
  $s_1$ & 0.5 & 0.5 \\ 
  $s_2$ & 0.2 & 0.8 \\ \hline
\end{tabular}
\end{center}
\end{minipage}
&
\begin{minipage}[ht]{11cm}
\begin{center}
\begin{tabular}{|c| c | c | c | c | c | c |} \hline
 $w$ & $t_1 $ & $t_2$ & $t_3$ & $t_4$ & $p(w)$ & $p(w|L_k)$\\
  & $(s_1)$&$(s_1)$&$(s_2)$& $(s_2)$ & & \\ \hline
$w_1$ &  $x$ & $x$ & $y$ & $y$ &  $0.5 \times 0.5 \times 0.8 \times
0.8 = 0.16$ &  $0.16/0.33 = 0.48$  \\ \hline
 $w_2$ & $x$ & $y$ & $x$ & $y$ &  $0.5 \times 0.5 \times 0.2 \times 0.8 = 0.04$ &  $0.04/0.33 = 0.12$ \\ \hline
$w_3$ &  $x$ & $y$ & $y$ & $x$ &  $0.5 \times 0.5 \times 0.8 \times
0.2 = 0.04$ &  $0.04/0.33 = 0.12$\\ \hline
  $w_4$ & $y$ & $x$ & $x$ & $y$ &  $0.5 \times 0.5
\times 0.2 \times 0.8 = 0.04$ &  $0.04/0.33 = 0.12$\\ \hline
 $w_5$ & $y$ & $x$ & $y$ & $x$ &  $0.5 \times 0.5 \times 0.8 \times 0.2 = 0.04$ &  $0.04/0.33 = 0.12$ \\ \hline
  $w_6$ &$y$ & $y$ & $x$ & $x$ &  $0.5 \times 0.5 \times 0.2 \times 0.2 = 0.01$ &  $0.01/0.33 = 0.03$ \\ \hline
\end{tabular}
\end{center}
\end{minipage}
\\
 (a) conditional distribution & (b) $p(w)$ and $p(w|L_k)$
\end{tabular}
\caption{An example illustrating the computation of $p(t_j:x)$}
\label{tab:exampleIllustrateProbLocalComputation} 
\end{table*}

\section{Probability Formulation}
\label{sec:probabilityFormulation}

For the sake of illustration, in this section, we consider a certain attribute set $\mathcal{A}$
and a sensitive value $x$. We will consider any attribute set
and any sensitive value in Section~\ref{sec:algDataPublishing}.

Suppose there are $m$ possible signatures for attribute set
$\mathcal{A}$, namely $s_1, s_2, ..., s_m$. Let $G$ be the
background knowledge consisting of the set of all QI based
distributions. In $G$, the probability that $s_i$ is linked to a
sensitive value $x$ is given by
$p(s_i:x )$.


Given $G$, the formula for $p(t:x)$, the probability that a tuple
$t$ is linked to sensitive value $x$, is derived below.

In the following, we consider the anonymized dataset $T^*$.
Suppose $t$ belongs to
A-group $L_k$ in $T^*$. For the ease of reference, let us summarize the notations
that we use in Table \ref{tab:notation}. We shall need the following
definitions.

\begin{table}[tbp]
\center  \small
\begin{tabular}{|c|l|}
  \hline
  $L_k$ & an A-group (anonymized group) in the \\ & anonymized dataset\\
  $\mathcal{A}$ & set of attributes
  e.g. \{``Nationality", ``Zipcode"\} \\
  $t_1,...,t_N$ & tuples in an $A$-group \\
  $s_1,...,s_m$ & signatures for $\mathcal{A}$, e.g.\{``American", ``55501"\}\\
  & multiple tuples $t_j$'s can map to the
  same $s_i$ \\
  $x,y$ &  sensitive values \\
  $p(t_j:x)$ & probability that tuple $t_j$ is linked to
  value $x$ \\
  $p(s_i:x)$ & probability that signature $s_i$ is linked to $x$ \\
  $f_i$ &  a simplified notation for $p(s_i:x)$ \\
  $w$ & a possible world: an assignment of the tuples\\
  & in A-group $L_k$ to the
  sensitive values in $L_k$ \\
$\mathcal{W}_k$ & set of all possible worlds $w$ for $L_k$\\
$\mathcal{W}_k^{(t_j:x)}$ &  set of all possible worlds $w$ in $\mathcal{W}_k$\\
&
in which $t_j$ is assigned value $x$\\
  $p(w|L_k)$ &  probability that $w$ occurs given A-group $L_k$
  \\
$p_{j,w}$ &
  the probability that $t_j$ is linked to a value in the \\
  & sensitive attribute as
 specified in $w$ \\
 \hline
  \end{tabular}
  \caption{Notations}
\label{tab:notation} 
\end{table}

\begin{definition}[Possible World]
Consider an A-group $L_k$ with $N$ tuples, namely $t_1, t_2, ...,
t_N$, with corresponding values in sensitive attribute $X$ of
$\gamma_{1}, \gamma_{2}, ... \gamma_{N}$. A possible world $w$ for
$L_k$ is a possible assignment mapping the tuples in set $\{t_1,
t_2, ..., t_N\}$ to values in multi-set $\{\gamma_{1}, \gamma_{2},
... \gamma_{N}\}$ in $L_k$.
\end{definition}

Given an A-group $L_k$ with a set of tuples and a multi-set of the
values in $X$. Considering all possible worlds, we form a sample
space. More precisely, the \textbf{sample space} $\Omega_{w|L_k}$
consists of all the possible assignments of the sensitive values in
$L_k$ to the $N$ tuples in $L_k$. For each such possible world $w$,
according to the QI based distribution $G$ based on attribute set
$\mathcal{A}$, we can determine the probability $p(w|L_k)$ that $w$
occurs given $L_k$.

\begin{definition}[Primitive Events, Projected Events]
A mapping $t:x$ from an individual or tuple $t$ to a value $x$ in
the set of sensitive attributes is called a \emph{primitive event}.
Suppose $t$ matches signature $s$. Let us call an event for the
corresponding signature, ``$s:x$", a \emph{projected event} for $t$.
Note that this projected event belongs to sample space $\Omega_s$.
\end{definition}

A primitive event is an event in the sample space $\Omega_{w|L_k}$.
The probability of such an event, $p(t:x)$, is the probability of
interest for the adversary. The probability of the projected event,
$p(s:x)$, is in the QI based distribution $G$.

Similar to \cite{l-diversity,XT06b,WFW+07}, we assume that the
linkage of a value in $X$ to an individual is independent of the
linkage of a value in $X$ to another individual. For example,
whether an American suffers from Heart Disease is independent of
whether a Japanese suffers from Heart Disease. Thus, for a possible
world $w$ for $L_k$, the probability that $w$ occurs given $L_k$ is
proportional to the product of the probabilities of the
corresponding projected events for the tuples $t_1, ... t_N$ in
$L_k$, we shall denote this product as $p(w)$:
\begin{eqnarray}
  p(w ) =  p_{1, w} \times p_{2, w} \times ... \times p_{N, w}
  \label{eqn1}
\end{eqnarray}

where $p_{j,w}$ is
  the probability that $t_j$ is linked to a value in the sensitive attribute
  specified in $w$.
Suppose $t_j$ matches signature $s_i$. If $t_j$ is linked to $x$ in
$w$, then $p_{j, w} = p(s_i:x )$.


 Let the
set of all the possible worlds for $L_k$ be $\mathcal{W}_k$. The sum
of probabilities of all the possible worlds given $L_k$ must be 1,
since they form the sample space $\Omega_{w|L_k}$. Hence, the
probability of $w$ given $L_k$ is given by:

For $w \in \mathcal{W}_k$, we have
\begin{eqnarray}
p(w|L_k) = \frac{ p(w)}{\mbox{$\sum_{w' \in \mathcal{W}_k}$}
p(w')} \label{eqn2}
\end{eqnarray}


Our objective is to find the probability that an individual $t_j$
 in $L_k$ is linked to a sensitive value $x$.
 This is given by the sum of the probabilities
 $p(w |L_k)$ of all the possible
 worlds $w$ where $t_j$ is linked to $x$.
\begin{eqnarray}
  p(t_j:x) = \mbox{$\sum_{w \in \mathcal{W}_k^{(t_j:x)}}$} \hspace*{0.2cm} p(w | L_k)
   \label{eqn-tupleLinkage}
\end{eqnarray}
where $\mathcal{W}_k^{(t_j:x)}$ is a set of all possible 
worlds $w$ in
$\mathcal{W}_k$ in which $t_j$ is assigned value $x$.

\begin{example} \em
Consider an A-group $L_k$ in a published table $T^*$. Suppose there
are four tuples, $t_1, t_2, t_3$ and $t_4$, with the $X$ values of
$x, x, y, y$ in $L_k$. Suppose the published table $T^*$ satisfies
2-diversity.

Consider the QI
based distribution $G$ based on a certain QI attribute set
$\mathcal{A}$ which contains two possible signatures $s_1$ and
$s_2$.
Table~\ref{tab:exampleIllustrateProbLocalComputation}(a) shows the
four global probabilities, namely $p(s_1:x)=0.5, p(s_1:y)=0.5,
p(s_2:x)=0.2, p(s_2,y)=0.8$.

Suppose $t_1, t_2, t_3$ and $t_4$ match signatures $s_1, s_1, s_2$
and $s_2$, respectively. There are six possible worlds $w$ as shown
in Table~\ref{tab:exampleIllustrateProbLocalComputation}(b). For
example, the first row is the possible world $w_1$ with mapping
\{$t_1:x$, $t_2:x$, $t_3:y$, $t_4:y$\}. The table also shows
the values $p(w)$ of the possible worlds. Take the first possible
world $w_1$ for illustration. From the QI based distribution in
Table~\ref{tab:exampleIllustrateProbLocalComputation}(a), $p(s_1:x)
= 0.5$ and $p(s_2: y) = 0.8$. Hence,
$p(w_1) = 0.5 \times 0.5 \times 0.8 \times 0.8 = 0.16$.
The sum of
$p(w)$ of all possible worlds from
Table~\ref{tab:exampleIllustrateProbLocalComputation}(b) is equal to
0.16 + 0.04 + 0.04 + 0.04 + 0.04 + 0.01 = 0.33. Consider
 $w_1$ again. Since $p(w_1) = 0.16$,
$p(w_1|L_k) = 0.16/0.33 = 0.48$.

Suppose the adversary is interested in the probability that $t_1$ is
linked to $x$. We obtain $p(t_1 : x )$ as follows. $w_1, w_2$ and
$w_3$, as shown in
Table~\ref{tab:exampleIllustrateProbLocalComputation}(b), contain
``$t_1:x$".
Thus, $p(t_1:x )$ is equal to the sum of the probabilities
$p(w_1|L_k), p(w_2|L_k)$ and $p(w_3|L_k)$.
$ p(t_1 : x ) 
= 0.48 + 0.12 + 0.12  = 0.72 $. Note that this is greater than 0.5,
the intended upper bound for 2-diversity that an individual is
linked to a sensitive value.\done
\end{example}

%% file: alg.tex
\section{Algorithm for Data Publishing}
\label{sec:algDataPublishing}

Given the formulation of $p(t:x)$, a naive approach for
$r$-robustness is to adopt some known anonymization algorithm $A$
and replace the probability measure in $A$ by $p(t:x)$. However, the
complexity of computing $p(t:x)$ is very high given the exponential
number of possible worlds. Moreover, $r$-robustness is not monotone
in the sense that an $A$-group that violates $r$-robustness may be
split into small groups that are $r$-robust, while known top-down
algorithms are based on monotone privacy conditions.

This section presents an algorithm for generating an $r$-robust
table that overcome the above problems.
Section~\ref{subsec:observation} first presents an important
theoretical property for this problem. Section~\ref{subsec:algASS}
then describes our proposed algorithm, ART.

\subsection{Theoretical Property}
\label{subsec:observation}

In Section~\ref{sec:intro}, we observe that privacy is breached
easily whenever an individual in an A-group has a much higher chance
of linking to a sensitive value compared with another individual in
the A-group. For example, consider the A-group $L_1$ in
Table~\ref{tab:genTable}. From the QI-based distribution
(Table~\ref{tab:globalDistMotivatingExample}), it is more likely
that American is linked to Heart Disease compared with Japanese, we
can deduce that Alex, an American, has Heart Disease with higher
probability. Note that the global probability of American linking to
Heart Disease, denoted by $f_1$, is 0.1 and the global probability
of Japanese linking to Heart Disease, denoted by $f_2$, is 0.003.
The difference in the global probabilities is $0.1 -0.003 = 0.097$.
Since the A-group size is small, the difference gives some
information to aid privacy breach. The difference in the global
probabilities and the A-group size are the \emph{properties} of the
A-group.

In the following, we have a theorem on the relationship between the
privacy guarantee and the properties of an A-group $L$.
Consider a tuple $t_v$ in an A-group $L$. We want to show that, if
the properties of $L$ satisfy some conditions, the privacy of $t_v$
can be guaranteed (i.e., $p(t_v:x) \le 1/r$). The conditions
essentially limits the deviations in the global probabilities in
terms of the group size.

In the following we consider the QI based distribution $G$ on a certain
attribute set $\mathcal{A}$. The algorithm to be described later
will consider multiple attribute sets.

\begin{definition}[Greatest Probability Deviation $\triangle$]
 Let $L$ be an A-group in $T^*$ with tuples $t_1,
t_2, ... t_N$ where $N$ is the group size and $N \ge r$. Let $x$ be
a sensitive value that appears once in $L$.
Without loss of generality, suppose tuple $t_v$ matches signature
$s_v$, $v \in [1, N]$. Thus, tuple $t_v$ has the QI based
probability (or global probability) linking to $x$ in $L$ equal to $p(s_v:x)$ = $f_v$.

Let $f_{max}$ be the greatest global probabilities in $L$ (i.e.,
$f_{max} = \max_{v \in [1, N]} f_v$). The probability deviation of
$t_v$ given $f_{max}$ is $\triangle_{v} = f_{max} - f_{v}$, $v \in [1,
N]$.
 \label{thmdef}
\end{definition}

Let us give some examples to illustrate the above notations. In our
running example of $L_1$, the group size $N$ is equal to 2. In
$L_1$, the first tuple (Alex) is $t_1$ and the second tuple  (Bob)
is $t_2$. Let $s_1$ = \{``American"\} and $s_2$ = \{``Japanese"\}.
Thus, $f_1 = 0.1$ and $f_2 = 0.003$. We know that $t_1$ matches
$s_1$ and $t_2$ matches $s_2$.
Since $f_{max}$ is the greatest global probabilities in $L$,
$f_{max}$ is equal to 0.1 (because  $f_1 = 0.1$ and  $f_2 = 0.003$).
Thus,  $\triangle_1 = f_{max} - f_1 = 0.1 - 0.1 =0$ and $\triangle_2
= f_{max} - f_2 = 0.1 - 0.003 = 0.097$.

\begin{theorem} \label{thm:triangleBound}
Let $r$ be the privacy parameter in $r$-robustness where $r > 1$.
Following the symbols in Definition \ref{thmdef},
%
%
if for all $v \in [1, N]$,
\begin{eqnarray}
 \label{triangle}
\triangle_{v} \le \frac{(N-r)f_{max}}{ f_{max}(r-1)/(1-f_{max}) +
(N-1)}
\end{eqnarray}
then for all $v \in [1, N]$, $p(t_{v}:x) \le 1/r$  
\end{theorem}
\textbf{Proof:} The proof is given in the appendix. \done

\begin{definition}[{$\triangle$, $\triangle_{max}$}]
$\triangle_{max}$ is defined to be the R.H.S. of Inequality~(\ref{triangle}).
That is,
$$ \triangle_{max} = \frac{(N-r)f_{max}}{ f_{max}(r-1)/(1-f_{max}) + (N-1) } $$
 $\mbox{Define \ \ \ \ }\triangle = \max_{v \in [1, N]} \{\triangle_v\}$ 
\end{definition}


 Hence,
$\triangle$ is the greatest difference in the global probabilities
linking to $x$ in an A-group. Note that $\triangle \ge 0$.
In our running example, since $\triangle_1 = 0$ and $\triangle_2 = 0.097$,
we have $\triangle = \max\{0, 0.097\} = 0.097.$

Consider another example. If an A-group $L$ contains three tuples
matching $s_1, s_2$ and $s_3$ with the global probabilities
$f_1=0.1$, $f_2=0.08$ and $f_3=0.09$. Then, $N = 3$ and $f_{max} =
0.1$. $\triangle = 0.1-0.08 = 0.02$. Suppose $r = 2$. The R.H.S. of
(\ref{triangle}) is $\triangle_{max} =(3-2) \times 0.1/[0.1\times
(2-1)/(1-0.1)+(3-1)] = 0.0474$. Since $\triangle < 0.0474$, from
Theorem~\ref{thm:triangleBound}, for all tuples $t_v$ in $L$,
$p(t_v:x) \le 1/r$ where $r = 2$.

Let us consider the effects of the values of $f_{max}$ and $N$ to
understand the physical meaning of Theorem~\ref{thm:triangleBound}.
 If $f_{max} = 1$ or $f_{max} = 0$, then
$\triangle \le 0$. Hence, the QI based distributions of all tuples in
$L$ should be the same to guarantee privacy.

\begin{table}
\begin{tabular}{c}
\begin{minipage}[ht]{8cm}
\center \scriptsize
\begin{tabular}{| c | c | c | c |} \hline
$N$ &  $r$ & $f_{max}$   & $\triangle_{max}$ \\ \hline 3 &  2 &  0.1 &
0.0474 \\ \hline 3 &  2 &  0.3 & 0.1235 \\ \hline 3 &  2 &  0.5 &
0.1667 \\ \hline
3 &  2 &  0.9 & 0.0818 \\ \hline
4 &  2 &  0.3 & 0.1750 \\ \hline 6 &  2 &  0.3 & 0.2211 \\ \hline 6
&  3 &  0.3 & 0.1537 \\ \hline 6 &  4 &  0.3 & 0.0955 \\ \hline
\end{tabular}
\caption{Values of $\triangle_{max}$ with some chosen values of $N,
r$ and $f_{max}$} \label{tab:triangleChosenValue} 
\end{minipage}
\end{tabular}
\end{table}

Table~\ref{tab:triangleChosenValue} shows the values of
$\triangle_{max}$ with some chosen values of $N, r$ and $f$. 
It can be seen that $\triangle_{max}$ is small when $f$ is near the
extreme values of 0 or 1, since the global probability of a tuple
is more pronounced.

Consider Inequality~(\ref{triangle}).
If $N \rightarrow \infty$, then $\triangle \le f_{max}$. Since
$f_{max}$ is the greatest possible global probability in $L$, it
means that $\triangle$ can be any feasible value (i.e., $0 \le
\triangle \le f_{max}$). Therefore, when the A-group is extremely
large, under Theorem~\ref{thm:triangleBound}, there will be no
privacy breach. When $N = r$, $\triangle \le 0$. That is, the global
probabilities of all tuples in $L$ should be equal. Otherwise,
there may be a privacy breach. Furthermore, $N$ has the following
relation with $\triangle_{max}$.


\begin{lemma}
\label{triangle2}
$\triangle_{max}$ is a monotonic increasing function on $N$.
\end{lemma}
%
%
\textbf{Proof:}Let $f = f_{max}$,
  $
  \frac{d \triangle_{max}}{d N} = \frac{(r - 1)
\times \frac{f^2}{1-f} +(r - 1)\times f}{[(r-1) \times \frac{f}{1-f}
+ (N-1)]^2}
 \ge 0$
\done

From the above, in order to guarantee $p(t_v:x) \le 1/r$,
we can increase the size $N$ of the A-group $L$. With a greater
value of $N$, the upper bound $\triangle_{max}$ increases, and the
constraint as dictated by Inequality~(\ref{triangle}) is relaxed,
making it easier to reach the guarantee.



\subsection{Algorithm ART}
\label{subsec:algASS}

Based on Theorem~\ref{thm:triangleBound}, we propose an
\underline{A}lgorithm generating $r$-\underline{R}obust
\underline{T}able called \emph{ART}.
If an A-group $L$ satisfies the inequality in
Theorem~\ref{thm:triangleBound} with respect to attribute set
$\mathcal{A}$ and, in $L$, each sensitive value occurs at most once,
we say that $L$ satisfies the \emph{QI based distribution bound
condition} with respect to $\mathcal{A}$. Otherwise, $L$ violates
the QI based distribution bound condition.

In the algorithm, initially, each individual forms an independent
A-group. The algorithm repeatedly looks for any A-group such that
there exists an attribute set $\mathcal{A}$ where it
violates the QI based distribution bound condition with respect to $\mathcal{A}$. Such a group is merged with other existing groups so
that the resulting group satisfies the condition.
After merging, the number of tuples in $L$, $N$, is increased.
Then, by Lemma \ref{triangle2}, $\triangle_{max}$
is also increased. The constraint by Inequality~(\ref{triangle}) is
relaxed and it is more likely to satisfy the QI based distribution
bound condition. When a final solution is reached, each individual
is linked to any sensitive value with probability at most $1/r$.

Specifically, algorithm ART
involves two major steps.

\begin{itemize}
  \item
  \textbf{Step 1} (Individual A-group Formation): For each tuple $t$ in the table $T$,
     we form an A-group $L$ containing $t$ only.
  \item \textbf{Step 2} (Merging): For \emph{each} sensitive value $x$, while there exists an A-group $L$ and
   an attribute set $\mathcal{A}$ such that $L$
violates the QI based distribution bound condition with respect to $\mathcal{A}$,
  we find a set $\mathcal{L}$ of A-groups such that, after merging all A-groups in $\mathcal{L}$
  with $L$, the merged A-group satisfies the QI based distribution bound condition with respect
  to any attribute set $\mathcal{A}$. 
\end{itemize}


%
The idea of Step 2 is to keep the $\triangle$ value in $L$ with
respect to $\mathcal{A}$ unchanged or only slightly increased after
merging.
At the same time, we also make sure that each merged A-group
contains at most one $x$ for any sensitive value $x$. 
Before going into the details of Step 2, we need to define a new
term. Given an A-group $L$, another A-group $L'$ is called a
\emph{closest} A-group with respect to $L$ if, after merging $L'$
and $L$, the increase in the value of $\triangle$ with respect to
\emph{any} attribute set is the smallest among all possible
A-groups.

\begin{definition}[Closest A-group]
Suppose $\triangle_{before, \mathcal{A}}$ represents $\triangle$
with respect to an attribute set $\mathcal{A}$ in $L$ and
$\triangle_{after, \mathcal{A}}(L, L')$ represents $\triangle$ with
respect to an attribute set $\mathcal{A}$ in the A-group obtained by
merging $L$ and $L'$.

Let $D_{\mathcal{A}}(L, L') = \triangle_{after, \mathcal{A}}(L, L')
- \triangle_{before,
\mathcal{A}}$. 

Let $D(L, L') = \sum_{\mathcal{A}}D_{\mathcal{A}}(L, L')$.

$L'$ is a \emph{closest} A-group with respect to $L$
if $D(L, L') =
 \min_{L''}\{D(L, L'')\}$. 
\end{definition}

We are ready to describe Step 2 in details. Let $Y(L)$ be the set of
sensitive values which appear in an A-group $L$. Given an A-group,
it is easy to derive $\triangle$ and $f_{max}$. Note that $r$ is a
user parameter.
After we know $\triangle$, $f_{max}$ and $r$, we can derive the
\emph{expected minimum size} of $L$ based on the QI based
distribution bound condition with respect to $\mathcal{A}$, denoted
by $N_o$.
By replacing $N$ with $N_o$ and
changing the subject of
Inequality~(\ref{triangle}) in the QI based distribution condition to $N_o$, we have
\begin{eqnarray}
N_o \ge \frac{(f_{max} (r - 1)\triangle)/(1-f_{max}) - \triangle + r
f_{max}}{(f_{max} - \triangle)} \label{ineq:howMany}
\end{eqnarray}
Let us choose a smallest integer $N_o'$ such that the above inequality holds.
We calculate $N_o'$ for \emph{every} attribute set $\mathcal{A}$
and choose the greatest values of $N_o'$ as our final $N_o'$.
If the total number of tuples in $L$, $N$, is smaller than $N_o'$,
then we have to choose additional $N_o'-N$ tuples to be merged with
$L$.
We choose a closest A-group $L'$ with respect to $L$ where $L'$ does
not contain any sensitive value in $Y(L)$.  $L'$ is merged with $L$,
and $\triangle$, $f$ and $N_o'$ are updated accordingly. 
%
If the updated $N$ value is still smaller than $N_o'$, then we
repeatedly continue the above process.

\begin{theorem}
Any table $T^*$ generated by Algorithm ART is $r$-robust. 
\end{theorem}

%% file: exp.tex
\section{Empirical Study}
\label{sec:exp}

%

A Pentium IV 2.2GHz PC with 1GB RAM was used to conduct our
experiment. The algorithm was implemented in C/C++. We adopted the
publicly available dataset, Adult Database, from the UCIrvine
Machine Learning Repository \cite{UCIrvine}. This dataset (5.5MB)
was also adopted by
\cite{l-diversity,WFW+07}.
We used a configuration similar to
\cite{l-diversity,WFW+07}. The records with unknown values
were first eliminated resulting in a dataset with 45,222 tuples
(5.4MB). Nine attributes were chosen in our experiment, namely Age,
Work Class, Marital Status, Occupation,
Race, Sex, Native Country, Salary Class and Education. 
By default, we chose the first five 
attributes and the last
attribute 
as the
quasi-identifer and the sensitive attribute, respectively.
Similar to \cite{WFW+07}, in attribute ``Education",
all values representing the education levels before ``secondary" (or
``9th-10th") such as ``1st-4th", ``5th-6th" and ``7th-8th" are
regarded as a sensitive value set where an adversary checks whether
each individual is linked to this set more than $1/r$, where $r$ is
a parameter.

There are 3.46\% tuples with education levels before ``secondary".
Since there is a set $\mathcal{G}$ of multiple QI based
distributions $G$, we can calculate $p(t:x)$ for different $G$'s and
different $x$'s. We take the greatest such value to report as the
probability that individual $t$ is linked to some sensitive value
since this corresponds to the worst case privacy breach.

We compared our proposed algorithm \emph{ART} with four algorithms,
\emph{Anatomy} \cite{XT06b}, \emph{MASK} \cite{WFW+07}, \emph{Injector} \cite{LL08}
and \emph{$t$-closeness} \cite{LL07}.
They are selected because they consider $l$-diversity or similar privacy
requirements, so we need
only set $l = r$. We are interested to know the overhead required in
our approach in order to achieve $r$-robustness. When we compared
\emph{ART} with \emph{Anatomy}, we set $l = r$. When we compared it
with \emph{MASK}, the parameters $k$ and $m$ used in MASK are set to
$r$. For \emph{Injector}, the parameters $minConf$, $minExp$ and $l$
are set to 1, 0.9 and $r$, respectively, which are
the default settings in \cite{LL08}. For \emph{$t$-closeness},
similar to \cite{LL07}, we set $t = 0.2$. We evaluate the algorithms in terms of four measurements:
(1) \emph{execution time}, (2) \emph{relative error ratio},
(3) the \emph{proportion of problematic tuples among all sensitive tuples} and
(4) the \emph{average value of $\triangle$}.

(1) \emph{Execution time:} We measured the execution time
of algorithms.
(2) \emph{Relative error ratio:}
 As in \cite{XT06b,WFW+07,LL08},
we measure the error by the \emph{relative error ratio} in answering
an aggregate query. We adopt both the form of the aggregate query and
the parameters of the \emph{query dimensionality} $qd$ and the
\emph{expected query selectivity} $s$ from \cite{XT06b,WFW+07,LL08}.
For each evaluation in the case of two anonymized tables, we
performed 10,000 queries and then reported the average relative
error ratio. By default, we set $s=0.05$ and $qd$ to be the QI
size.
(3) \emph{Proportion of problematic tuples among all sensitive tuples:}
According to the probability formulation in Section~\ref{sec:probabilityFormulation},
according to the anonymized table generated by all algorithms,
we can calculate the probability that a tuple is linked to
a sensitive value set.
If the tuple has the probability $ > 1/r$, it is said to be a \emph{problematic
tuple}.
The tuples linking
to sensitive values in the original table are called \emph{sensitive tuples}.
In our experiments, we measure
the proportion of problematic tuples among all sensitive tuples.
(4) \emph{Average value of $\triangle$:}
More formally, the average value of $\triangle$ is evaluated with
respect to every attribute set $\mathcal{A}$ containing large
samples. Consider a sensitive value $x$. With respect to a certain
attribute set $\mathcal{A}$, the average value of $\triangle$
denoted by $H_{\mathcal{A}}$ is equal to $\mbox{$\frac{1}{u}$}
\sum_{L \in T^*}\triangle_L$, where $u$ is the total number of
A-groups in $T^*$ and $\triangle_L$ is the greatest difference in
the global probability linking to a sensitive value $x$ with respect
to $\mathcal{A}$ in an A-group $L$. Let $B$ be the set of all
attribute sets $\mathcal{A}$ containing large samples. With respect
to every attribute set in $B$, the average value of $\triangle$ is
equal to $\mbox{$\frac{1}{|B|}$}\sum_{\mathcal{A} \in
B}H_{\mathcal{A}}$. We perform the same steps for every sensitive
value $x$ and take the average as the reporting average value of
$\triangle$. For each measurement, we conducted the experiments 100
times and took the average.

We conducted the experiments by varying four factors: (1)
 the QI size, (2) $r$, 
(3) query dimensionality $qd$ 
and (4) selectivity $s$. 

\begin{figure}[tb] 
\begin{tabular}{c c}
    \begin{minipage}[htbp]{4.0cm}
        \includegraphics[width=4.0cm,height=3.0cm]{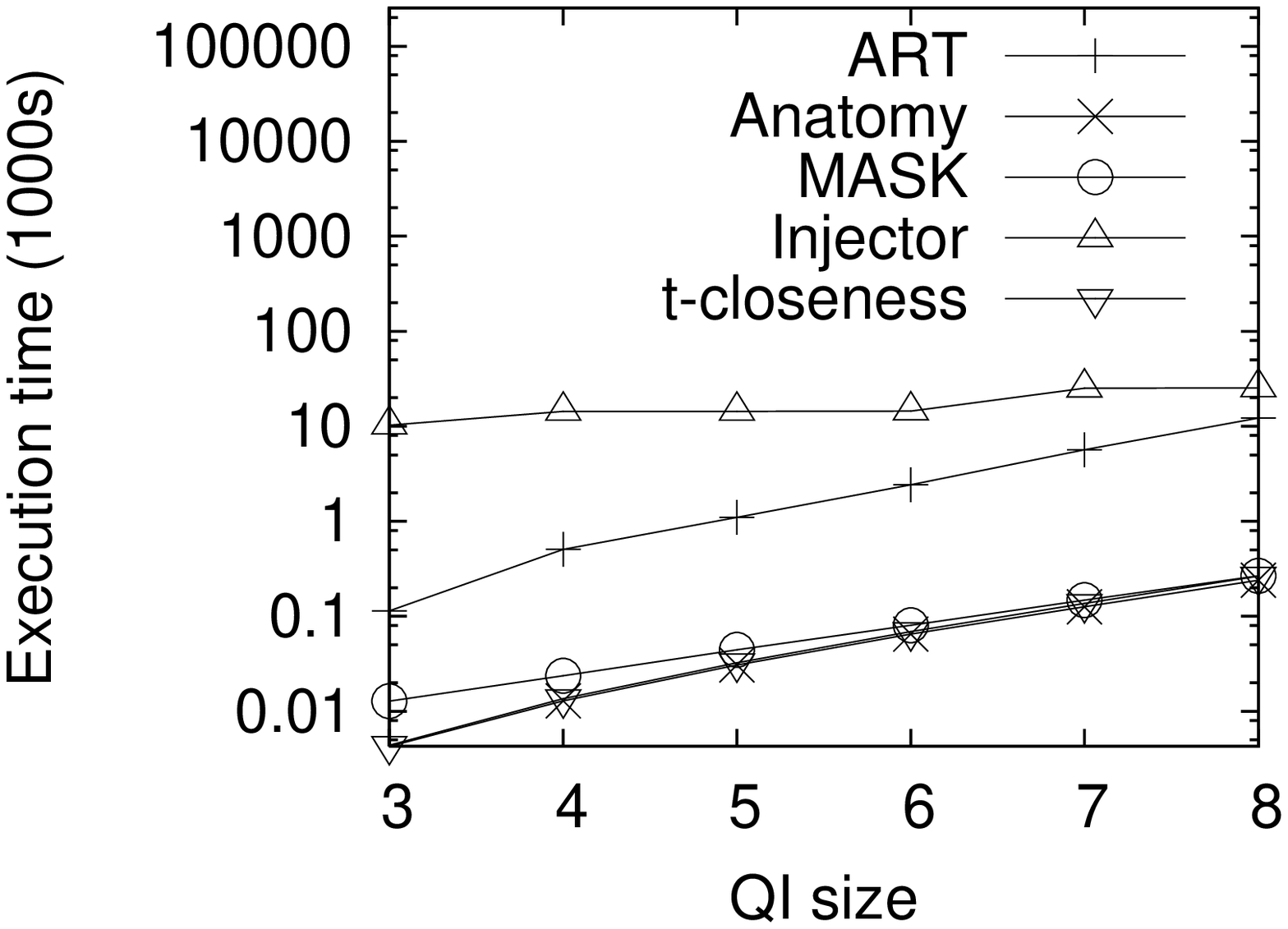}
    \end{minipage}
&
    \begin{minipage}[htbp]{4.0cm}
        \includegraphics[width=4.0cm,height=3.0cm]{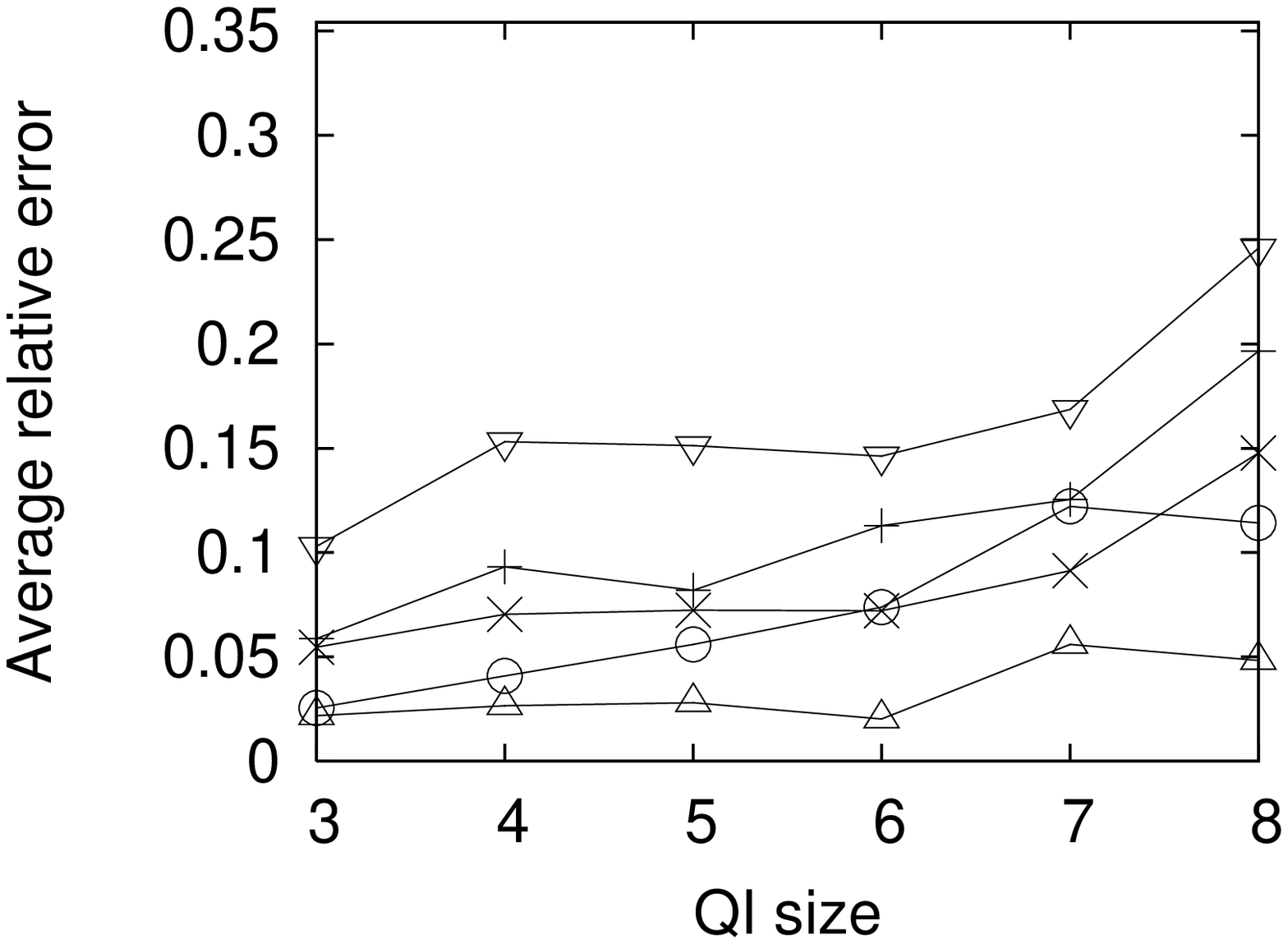}
    \end{minipage}
\\
(a)
&
(b)
\\
    \begin{minipage}[htbp]{4.0cm}
        \includegraphics[width=4.0cm,height=3.0cm]{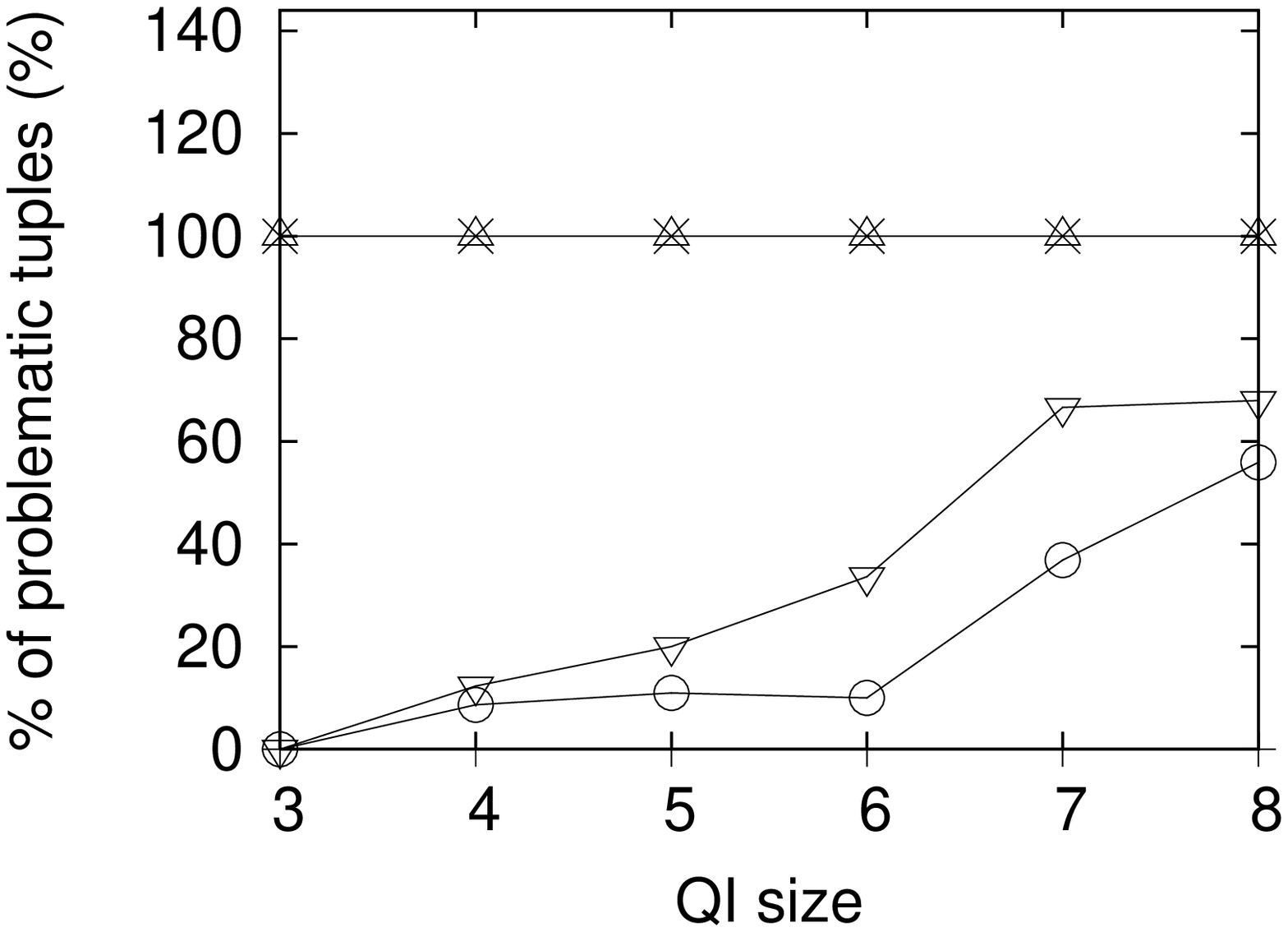}
    \end{minipage}
&
    \begin{minipage}[htbp]{4.0cm}
        \includegraphics[width=4.0cm,height=3.0cm]{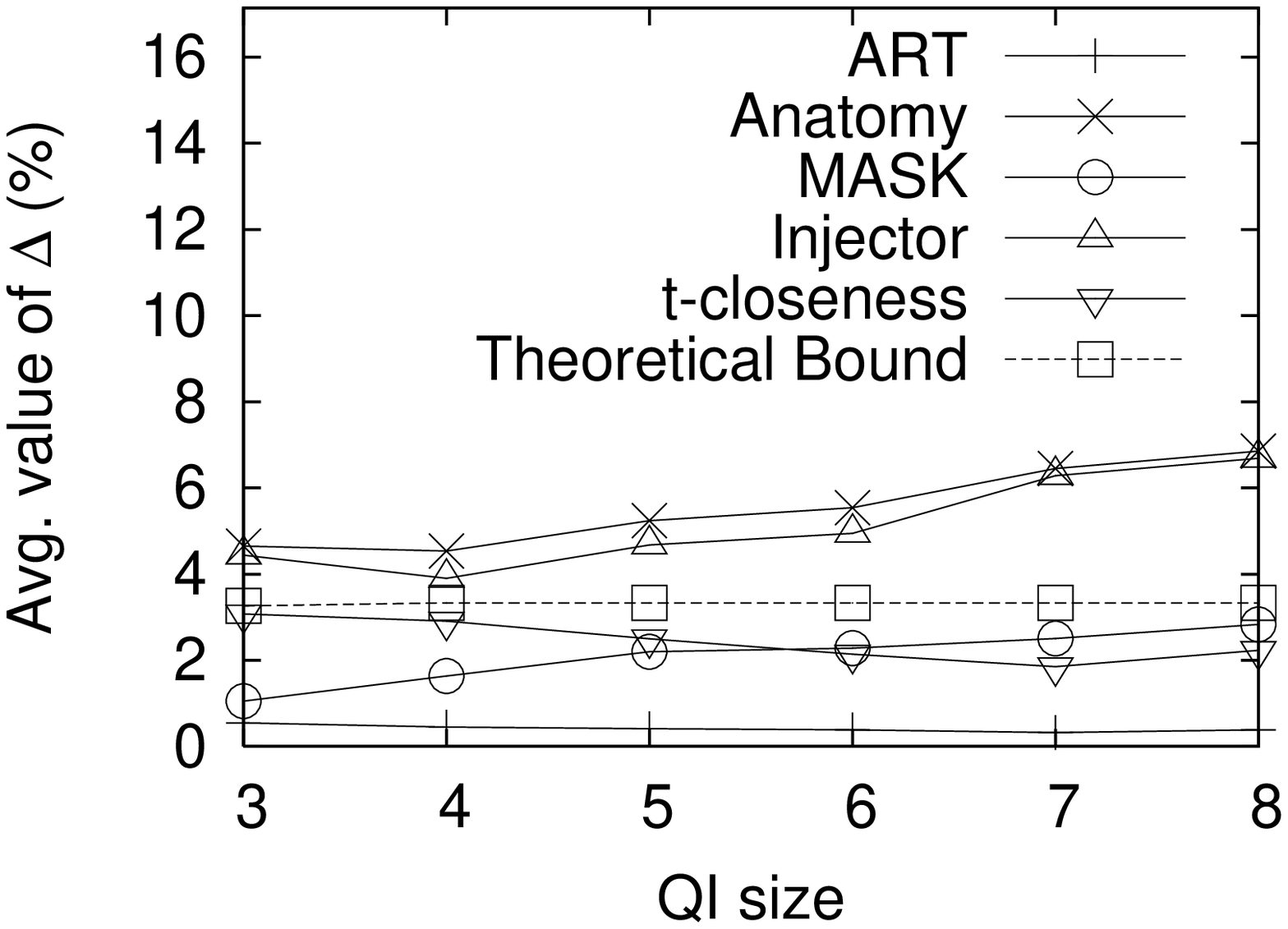}
    \end{minipage}
    \\
(c)
&
(d)
\end{tabular}
\caption{Effect of QI size ($r = 10$)}\label{fig:graphAgainstQID-m10}
\end{figure}

Figure~\ref{fig:graphAgainstQID-m10} shows the
results when $r$ is
set to 10. 
Figure~\ref{fig:graphAgainstQID-m10}(a) shows that
the execution time increases with the QI size
because the algorithms have to process more
QI attributes. \emph{ART} performs slower
compared with \emph{Anatomy}, \emph{MASK} and \emph{$t$-closeness}.
Since \emph{ART} requires to
compute the QI based distribution with respect
to every attribute set, when the QI size
increases, the increase in the execution time of \emph{ART}
is larger.


Figure~\ref{fig:graphAgainstQID-m10}(b) shows
that there is an increase in average relative error
when the QI size increases because it is more
difficult to form A-groups where
the difference in QI based distributions among
all tuples in an A-group is small when the QI size is larger.
Since \emph{$t$-closeness} is
a global recoding and causes a lot of unnecessary
generalizations, the average relative error is the
largest. Since \emph{Injector} tries to exclude
some sensitive values in an A-group, its
relative error is also small.

Figure~\ref{fig:graphAgainstQID-m10}(c) shows that the proportion
of problematic tuples among sensitive tuples increases with QI size.
With a larger QI size, there is a higher chance that
individual privacy breaches due to more attributes which
can be used to construct the QI based distributions.
\emph{MASK} has fewer privacy breaches compared with
\emph{$t$-closeness},
\emph{Anatomy} and \emph{Injector} because the side-effect of the minimization
of QI values in each A-group adopted in \emph{MASK}
makes the difference in the QI based distribution among
all tuples in each A-group smaller. Thus, the number of
individual with privacy breaches is smaller.
It is noted that there is no violation in \emph{ART}.

In Figure~\ref{fig:graphAgainstQID-m10}(d), we include
the theoretical bound of $\triangle_{max}$ from Theorem~\ref{thm:triangleBound}
for comparison. We use the bound of \emph{ART} as this theoretical bound because,
compared with
\emph{Anatomy} 
and \emph{Injector},
the size of A-groups formed in \emph{ART} is largest (which yields the largest bound).
Since the average value of $\triangle$ of \emph{Anatomy} 
and \emph{Injector} are greater than
this bound,
they may have privacy breaches as shown in Figure~\ref{fig:graphAgainstQID-m10}(c).
When the QI size increases, the average value of $\triangle$ with respect to every attribute set increases, as shown in
Figure~\ref{fig:graphAgainstQID-m10}(d). With a larger QI size,
during forming an A-group, we have to consider $\triangle$
with respect to more attribute sets. Thus,
it is more likely that an A-group has a larger average value of $\triangle$
with respect to every attribute set.
The average value of $\triangle$ is the largest in \emph{Anatomy} and \emph{Injector}, and
the next two largest in \emph{MASK} and \emph{$t$-closeness}. This is because
\emph{Anatomy} and \emph{Injector} does not take our QI based distribution directly into
the consideration for merging but \emph{MASK} and \emph{$t$-closeness} do indirectly
during the minimization of QI values.
In Figure~\ref{fig:graphAgainstQID-m10}(d), although
the average value of $\triangle$ of \emph{MASK}
is smaller than the theoretical bound of $\triangle$,
it is possible to breach privacy as shown in
 Figure~\ref{fig:graphAgainstQID-m10}(c) because this evaluation
 only shows the \emph{average} value and the actual $\triangle$  in some A-groups is larger than this bound.

We also conducted experiments when $r = 2$.
For the sake of space, we did not show the figures.
The results are also similar. But, the execution
time and 
the average
relative error
are smaller.
Since $r$ is smaller and thus $1/r$ is larger, the
average value of $\triangle$ is larger
when $r = 2$.

%% file: related.tex
\section{Related Work}
\label{sec:related}

With respect to attribute types considered for
data anonymization, there are two branches of studying. The first
branch is anonymization according to the QI attributes. A typical
model is $k$-anonymity
\cite{AggarwalICDT05}. 
The other branch is the consideration of both QI
attributes and sensitive attributes. Some examples
are 
\cite{l-diversity}, \cite{WLFW-kdd06}, 
\cite{LL07},
\cite{LL08}
and \cite{BS08}. In this
paper, we focus on this branch. We want to check whether the
probability that each individual is linked to any sensitive value is
at most a given threshold.

%
$l$-diversity \cite{l-diversity} proposes a model where $l$ is a
positive integer and each A-group contains $l$ ``well-represented"
values in the sensitive attribute. For $t$-closeness \cite{LL07}, the distribution in
each A-group in $T^*$ with respect to the sensitive attribute is
roughly equal to the distribution of the entire table $T^*$.

In the literature, different kinds of background knowledge are
considered \cite{l-diversity,MKM+07,WFW+07,LLZ09,GKS08,LL08,APZ06}.
\cite{MKM+07} considers another background knowledge
in form of implications.
\cite{WFW+07} discovers that the minimality principle of the
anonymization algorithm can also be used as background knowledge.
\cite{LLZ09} proposes to use the kernel estimation method
to mine the background knowledge from the original table.

\cite{LL08} finds that association rules can be mined from the
original table and thus can be used for
privacy protection during anonymization. 
%
In \cite{APZ06}, the problem of privacy attack by adversarial
association rule mining is investigated.
However, as pointed out in \cite{SMB97}, association rules used in
\cite{LL08} and \cite{APZ06} can contradict the true statistical
properties. 
Also the solution in \cite{APZ06} is to invalidate the rules, but
this will violate the data mining objectives of data publication.

%% file: concl.tex
\section{Conclusion}
\label{sec:concl}

In this paper, we consider
the worst-case
QI based distribution for privacy-preserving
data publishing.
Then,
we derive a theoretical property and propose
an algorithm which generates a dataset protecting
individual privacy in the presence of the worst-case
QI based distribution.
    Finally,
      we conducted experiments
      to show that our proposed algorithm is efficient and
      incurs low information loss.
For future work, we plan to investigate
how to anonymize the dataset with other kinds of
background knowledge that may be possessed by
the adversary.

%% file: proof.tex
\medskip

\section{Appendix}

\small

Here we prove our main theorem. Let us recap a few notations.

\bigskip

\begin{tabular}{|c|l|}
  \hline
  $p(s_i:x)$ & probability that signature $s_i$ is linked to $x$ \\
  $f_i$ &  a simplified notation for $p(s_i:x)$ \\
  $f$ & $f_{max}$, maximum $f_i$ value among all $i$'s\\
  \hline
\end{tabular}

\bigskip

\noindent\textbf{Proof of Theorem~\ref{thm:triangleBound}:}
%
Let $t_u$ be a tuple in $L$ with the greatest global probability
linking to $x$ in $L$ (i.e., for all tuples $t_v$ in $L$, $f_u \ge f_v$).
Besides, $f = f_u$.

Consider the set $W_u$ of  possible worlds where ``$t_u:x$" occurs.
Let $t_v$ be a tuple such that $\triangle_v = \max_{a \in [1,
N]}\{\triangle_a\}$. Consider the set of possible worlds $W_v$ where
``$t_v:x$" occurs.

Consider also the set of possible worlds $W_a$ where ``$t_a:x$"
occurs for an arbitrary $t_a$ where
$t_a \neq t_v$. We first want to show that 
 $p({W}_a) \ge p({W}_v)$, where $p({W}_a)$ is the probability that
 any world in $W_a$ occurs.
%

\begin{lemma}
For $a \in [1, N]$,  $p({W}_a) \ge p({W}_v)$.
\label{lemma:probInequalityRAndV}
\end{lemma}
\textbf{Proof of Lemma~\ref{lemma:probInequalityRAndV}:}
Since $\triangle_v = \max_{a \in [1, N]}\{\triangle_a\}$, $f_v \le
f_a$ and $(1-f_v) \ge (1-f_a)$. Hence,
\begin{eqnarray}
  f_{a}(1 - f_{v}) \ge f_{v}(1 - f_{a})
  \label{eqn:ineqaulityRTermVTerm}
\end{eqnarray}

For a world $w_v \in W_v$, $p({w}_v) = p_{1, {w}_v} \times ...
\times p_{N, {w}_v}$.

For a world $w_a \in W_a$, $p({w}_a) = p_{1, {w}_a} \times ...
\times p_{N, {w}_a}$.

Note that $p_{v, {w}_v} = f_{v}$ and $p_{a, {w}_a} = f_{a}$.

Since there is only one $x$ occurrence in $L$, $t_v$ is not assigned
with $x$ in any $w_a \in W_a$. Let $W'_a$ be a maximal subset of
$W_a$ where $t_v$'s are assigned to distinct $X$ values.
Obviously $\sum_{w_a \in W'_a} p_{v, {w}_a} = 1 - f_{v}$. 
Hence,
\begin{eqnarray}
  \mbox{$\sum_{w_a \in W'_a} (p_{a, {w}_a} \times p_{v, {w}_a}) = f_{a}(1 -
  f_{v})$}
  \label{eqn:firstRTerm}
\end{eqnarray}
Similarly, since $t_a$ is not assigned with $x$ in any $w_v \in
W_v$, we can find a maximal subset $W'_v$ in $W_v$ where $t_v$'s are
assigned to distinct $X$ values.
we have $\sum_{w_v \in W'_v} p_{a, {w}_v} = 1 - f_{a}$. 
\begin{eqnarray}
   \mbox{$\sum_{w_v \in W'_v} p_{v, {w}_v} \times p_{a, {w}_v} = f_{v}(1 -
   f_{a})$}
  \label{eqn:firstVTerm}
\end{eqnarray}
From (\ref{eqn:ineqaulityRTermVTerm}), (\ref{eqn:firstRTerm}), and
(\ref{eqn:firstVTerm}),
\begin{eqnarray}
  \mbox{$\sum_{w_a \in W'_a} p_{a, {w}_a} \times p_{v, {w}_a}
  \ge \sum_{w_v \in W'_v} p_{v, {w}_v} \times p_{a, {w}_v}$}
 \label{eqn:inequalityRTermVTermSubstitute}
\end{eqnarray}

For each $w_a \in W'_a$ we can find a unique $w_v$ in $W'_v$, so
that $f_v$ in $w_a$ and $f_a$ in $W_v$ are assigned the same
sensitive value.We say that $w_a$ and $w_v$ are matching.
 Let us further restrict $W'_v$ based on $W'_a$ in
such a way that the matching world $w_v$ in $W'_v$ for $w_a$ in
$W'_a$ has the same sensitive value assignments for the remaining
tuples. It is obvious that we can always form such an $W'_v$ from
and any $W_a$.
For matching $w_a$ and $w_v$,
\begin{eqnarray}
\mbox{$\prod_{i \not\in \{a,v\}} p_{i,w_a} = \prod_{i \not\in
\{a,v\}}
 p_{i,w_v}$}
  \label{eqn:equalityRTermVTerm}
\end{eqnarray}

Furthermore, $W_a$ can be partitioned into $W'_a$'s. and the union
of the corresponding $W'_v$ is equal to $W_v$.

From (\ref{eqn:inequalityRTermVTermSubstitute}) and
(\ref{eqn:equalityRTermVTerm}), we conclude that \[\mbox{$\sum_{w_a
\in W_a} p_{1, {w}_a} \times ... \times p_{N, {w}_a} \ge \sum_{w_v
\in W_v} p_{1, {w}_v} \times ... \times p_{N, {w}_v}$}\]
That is, $\sum_{w_a \in W_a} p(w_a) \geq \sum_{w_v \in W_v} p(w_v)$.

Therefore, for $a \in [1, N]$,
\begin{eqnarray}
  p({W}_a) \ge p({W}_v)
  \label{eqn:inequalityWorldRTermWorldVTerm}
\end{eqnarray}
This completes the proof of Lemma~\ref{lemma:probInequalityRAndV}.\done


\begin{lemma}
If $p(t_u:x) \le 1/r$, then $p(t_a:x) \le 1/r$ for all $a \in [1,
N]$. \label{lemma:oneHighestEnoughForProtection}
\end{lemma}
\textbf{Proof of Lemma~\ref{lemma:oneHighestEnoughForProtection}:}
By similar techniques used in the proof of
Lemma~\ref{lemma:probInequalityRAndV}, since $f_u \ge f_a$ for all
$a \in [1, N]$, we derive that $p(W_u) \ge p(W_a)$. Let $K =
\sum_{w' \in \mathcal{W}}p(w')$ where $\mathcal{W}$ is a set of all
possible worlds. Since $p(t_u:x) = p(W_u|L) = p(W_u)/K$ and
$p(t_a:x) = p(W_a|L) = p(W_a)/K$, we have $p(t_u:x)\ge p(t_a:x)$.
Thus,  if $p(t_u:x) \le 1/r$, then, for all $a \in [1, N]$,
$p(t_a:x) \le 1/r$.

This completes the proof of Lemma~\ref{lemma:oneHighestEnoughForProtection}.\done

Lemma~\ref{lemma:oneHighestEnoughForProtection} suggest that, once
$p(t_u:x)$ is bounded $1/r$, all other probabilities $p(t_a:x)$ in
the A-group are also bounded. In the following, we focus on
analyzing $p(t_u:x)$ only (instead of all probabilities $p(t_a:x)$).

Consider $p(t_u:x)$, which is equal to $p({w}_u|L)$. Let
$\mathcal{W}$ be a set of all possible worlds. Let
$\mathcal{W}^{(t_u:x)}$ be the set of all possible worlds with
``$t_u:x$". By definition $\mathcal{W}^{(t_u:x)} = W_u $ and there
are $N$ such sets of worlds in $\mathcal{W}$. Also,
\begin{eqnarray*}
p(W_u|L) & = & \frac{\sum_{w \in \mathcal{W}^{(t_u:x)}}p({w})}{\sum_{w \in \mathcal{W}}p({w})} \\
& = & \frac{\sum_{w \in \mathcal{W}^{(t_u:x)}}p({w})}{\sum_{w \in \mathcal{W}^{(t_u:x)}}p({w}) + \sum_{w' \in \mathcal{W}/\mathcal{W}^{(t_u:x)}}p({w}')}\\
& = & \frac{p({W}_u)}{p({W}_u) + \sum_{a \neq u}p({W}_a)}
\end{eqnarray*}

By Lemma~\ref{lemma:probInequalityRAndV}, $\sum_{a \neq u}p({W}_a)
\ge (N-1)p({W}_v)$. Hence,
\begin{eqnarray}
p(W_u | L) \le \frac{p({W}_u)}{p({W}_u) + (N-1)p({W}_v)}
 \label{eqn:inequalityProbWU}
\end{eqnarray}
%
From the proof of Lemma \ref{lemma:probInequalityRAndV}, $W_u$ and
$W_v$ can be partitioned into matching pairs of $W'_u$ and $W'_v$
where
  $\sum_{w_u \in W'_u} p(w_u) = f_u(1 - f_v) C$ for some $C$
  and
    $\sum_{w_v \in W'_v} p(w_v) = f_v(1 - f_u) C$.\\

 Therefore, we can
simplify Inequality~(\ref{eqn:inequalityProbWU}) as follows.
\begin{eqnarray}
p(w_u | L) \le \frac{f_u(1-f_v)}{f_u(1 - f_v) + (N-1)\times f_v(1 -
f_u)}
 \label{eqn:inequalityProbWU2}
\end{eqnarray}

Consider the term 
 $(N-1) \times f_v( 1 - f_u)$
in Inequality~(\ref{eqn:inequalityProbWU2})
\begin{eqnarray*}
 &  & (N-1) (1 - f_{u}) f_{v}\\
  & = & (N-1) (1 - f) (f - \triangle_v)\\
  & = & (r - 1) f (1 - f + \triangle_v) \times \frac{ (N-1) (1 - f) (f - \triangle_v)}{(r - 1) f (1 - f + \triangle_v)} \\
  & = & (r - 1) f_u( 1- f_v) \times \frac{ (N-1) (1 - f) }{(r - 1) f (\frac{1}{f - \triangle_v} - 1)}
\end{eqnarray*}

After substituting
$\triangle_{v} \le (N-r)f/[ \frac{f(r-1)}{1-f} + (N-1)]$ into the
above equation, with simple derivations, we obtain $$(N-1)\times
f_v(1 - f_u) \ge (r - 1) \times f_u(1 - f_v)$$

 With the above inequality,
Inequality~(\ref{eqn:inequalityProbWU2}) becomes
\begin{eqnarray}
p(W_u | L) \le 1/r 
 \label{eqn:inequalityProbWU3}
\end{eqnarray}
This completes the
proof of Theorem~\ref{thm:triangleBound}. \done